\documentclass{iaus}
\usepackage{graphicx}

\title[Microlensing towards LMC and M31]{Microlensing towards LMC and M31}

\author[Philippe Jetzer, Alain Milsztajn and Patrick Tisserand]%
{Ph. Jetzer$^1$, A. Milsztajn$^2$ and P. Tisserand$^2$}

\affiliation{$^1$Institute of Theoretical Physics, University of Z\"urich,
Winterthurerstasse 190, CH-8057 Z\"urich, Switzerland,
email: jetzer@physik.unizh.ch\\[\affilskip]
$^2$Dapnia, Service de Physique des Particules, CEA-Saclay, F-91191
 
Gif-sur-Yvette, France; email: amilsztajn@cea.fr, ptisserand@cea.fr}

\pubyear{2004}
\volume{225}
\pagerange{1--12}
\date{?? and in revised form ??}
\jname{Impact of Gravitational Lensing on Cosmology}
\editors{Mellier, Y. \& Meylan,G. eds.}
\begin{document}

\maketitle

\begin{abstract}
The nature and the location of the lenses discovered in the
microlensing surveys done so far towards the LMC remain unclear.

This contribution is comprised of two distinct parts. In the first part,
motivated by these questions, we compute the optical depth 
for the different intervening populations and
the number of expected events for self-lensing, using a recently drawn 
coherent picture of the geometrical structure and dynamics of the
LMC disk. By comparing the
theoretical quantities with the values of the observed events it
is possible to put some constraints on the location and the nature
of the machos. 
Clearly, given the large uncertainties and the few
events at our disposal it is not yet possible to draw sharp conclusions,
nevertheless we find that up to 
3-4 macho events might be due to lenses in LMC, which are most probably
low mass stars, but that hardly all events can be due to
self-lensing. A
plausible solution is that the events observed so far are due to
lenses belonging to different intervening populations: low mass
stars in the LMC, in the thick disk, in the spheroid and some true
machos in the halo of the Milky Way and the LMC itself.
We report also on
recent results 
of microlensing searches in direction of the M31 galaxy,
by using the pixel method. The present analysis still
does not allow yet to draw sharp conclusions on the macho content of the
M31 galaxy.

In the second part (section 5),
a preliminary account of the final results from the
EROS-2 programme is presented. Based on the analysis of
33 million LMC and SMC stars followed during 6.7 years, 
strict limits on the macho content of the galactic
halo are presented; they cover the range of macho masses between 0.0001 and
100 solar mass.  The limits are better than 20\% (resp. 5\%)
of the standard halo for masses between 0.0002 and 10 
(resp. 0.001 to 0.1) solar mass. 
This is presently the data set with the largest
sensitivity to halo machos. 
\end{abstract}

\section{Introduction}

Since Paczy\'nski's original proposal (\cite{pacz})
gravitational microlensing has been proven to be
a powerful tool for the detection of the dark matter
component in galactic haloes in the form of machos.
Searches in our Galaxy towards the LMC show
that up to 20\% of the halo could be formed
by objects of around $M \sim 0.4\,M_\odot$.

However, the location and the nature of the microlensing events found so
far towards the LMC is still a matter of
controversy. The MACHO collaboration found 13 to 17 events in 5.7
years of observations, with a mass for the lenses estimated to be
in the range $0.15 - 0.9 ~M_{\odot}$  assuming a
standard spherical Galactic halo (\cite{alcock00a}). From this,
they derived a total optical depth towards the LMC
of $\tau= 1.2^{+0.4}_{-0.3} \times 10^{-7}$. 
The EROS2 collaboration has announced at this symposium the detection
of 4 candidate events (one of which being most probably due to a
lens located in the disk of our Galaxy) based on 6.7 years of observation 
but monitoring about three times as much stars as the MACHO collaboration.
These EROS2 results are detailed in section 5 of this contribution.
The MACHO collaboration monitored primarily the
central part of the LMC, whereas the EROS2 experiment covered a
larger solid angle but in less crowded fields. The
EROS2 microlensing rate should thus be less affected by
self-lensing. This might be the reason for the fewer events seen
by EROS2.

The analysis of \cite{jetzer02} and \cite{mancini} 
has shown that probably the observed
events are distributed among different galactic components (disk,
spheroid, galactic halo, LMC halo and self-lensing). This means that
the lenses do not belong all to the same population and their
astrophysical features can differ deeply.

Some of the events found by the MACHO team are most probably due
to self-lensing: the event MACHO-LMC-9 is a double lens with
caustic crossing (\cite{alcock00b}) and its proper motion is
very low, thus favouring an interpretation as a double lens within
the LMC. The source star for the event MACHO-LMC-14 is double
and this has allowed to conclude that the
lens is most probably in the LMC. 
The event LMC-5 is due to a disk lens and indeed
the lens has even been observed with the HST, see \cite{alc01nat}. 
The other stars which have been
microlensed were also observed but no lens could be detected, thus
implying that the lens cannot be a disk star but has to be either
a true halo object or a faint star or brown dwarf in the LMC
itself.

Thus up to now the question of the location of the observed MACHO events
is unsolved and still subject to discussion. Clearly, with much more
events at disposal one might solve this problem
by looking for instance at their spatial distribution. 
To this end a correct knowledge of the structure
and dynamics of the luminous part of the LMC is essential, and 
we take advantage of a new picture of the LMC disk.

Searches towards M31
have also been proposed (\cite{crotts92,agape93,jetzer94}).
This allows to probe a different line of sight in our Galaxy,
to globally test the M31 halo and, furthermore,
the high inclination of the M31 disk is expected
to provide a strong signature (spatial distribution) for 
halo microlensing signals.
For extragalactic targets, due to the distance, 
the sources for microlensing signals
are not resolved. This calls for an original technique, 
the \emph{pixel method}, 
the detection of flux variations of unresolved 
sources,
the main point being that one follows flux variations of every pixel
in the image instead of single stars. 
Several collaborations have recently 
presented a handful of microlensing events
(SLOTT-AGAPE \cite{mdm2},
POINT-AGAPE \cite{point03},
WeCapp \cite{wecapp03} 
and MEGA \cite{mega}).

\section{LMC optical depth} \label{tau}

In a series of 
interesting papers (\cite{marel01a,marel02}), a
new coherent picture of the geometrical structure and dynamics of
the LMC has been given. In the following analysis we use
this model 
and adopt the same coordinate system and notations as in van der
Marel et al. The
center of the disk coincides with the center of the bar 
and its distance from us is $D_{0} = 50.1 \pm
2.5 \, \mathrm{kpc}$. We take a bar mass
$M_{\mathrm{bar}}=1/5\,M_{\mathrm{disk}}$ with
$M_{\mathrm{bar}}+M_{\mathrm{disk}}=M_{\mathrm{vis}}=2.7 \times
10^{9}\, \mathrm{M}_{\odot}$.

We consider also the LMC halo contribution to the optical depth.
We use 
two different models to describe the halo density profile:
a spherical halo and an ellipsoidal halo. The
values of the parameters have been chosen so that the models have
roughly the same mass within the same radius. For the spherical model
we adopt  a classical pseudo-isothermal spherical  density
profile, with a halo mass within a radius of 8.9
kpc  equal to $5.5\times10^{9}~\mathrm{M}_{\odot}$.

\begin{figure}
\centering
\resizebox{12cm}{!}{\includegraphics{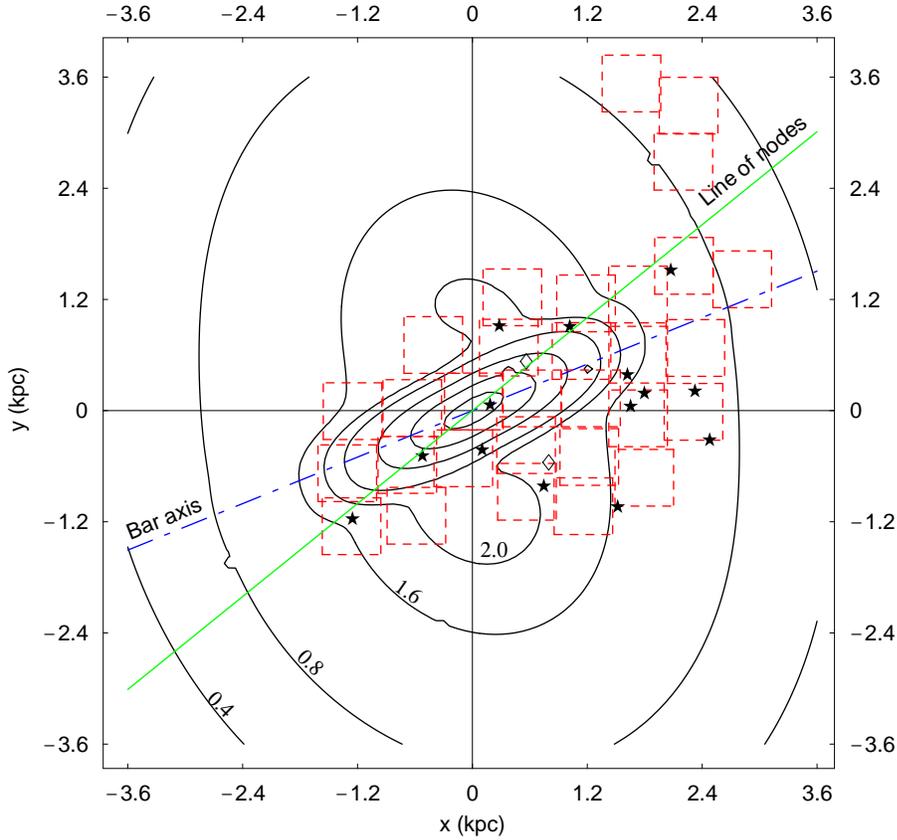}}
\caption{Contour map of the optical depth for self-lensing.  The
 locations of the MACHO fields and of their microlensing candidates
 are also shown. The numerical values are in $10^{-8}$ units.
 The innermost contours correspond to
 values $2.4\times 10^{-8}$, $3.2\times 10^{-8}$, $4.0\times 10^{-8}$
 and $4.6\times 10^{-8}$ respectively.}
\label{SL}
\end{figure}

The computation of the optical depth is made following 
the method outlined in \cite{jetzer02}.
In Fig. \ref{SL} we report the optical depth contour map for
self-lensing, i.e. for events where both the sources and the
lenses belong to the disk and/or to the bulge of LMC. As expected,
there is almost no near-far asymmetry
and the maximum value of the optical depth, $\tau_{\mathrm{max}}
\simeq 4.80 \times 10^{-8}$, is reached in the center of LMC. The
optical depth then  rapidly decreases, when moving, for instance,
along a line going through the center and perpendicular to the
minor axis of the elliptical disk, that coincides also with the
major axis of the bar. In a range of about only $0.80
\,\mathrm{kpc}$ the optical depth  quickly falls to $\tau \simeq
2 \times 10^{-8}$, and afterwards it decreases slowly to lower
values.
We computed also the optical depth
contour maps for lenses belonging to the halo of LMC
in the case of spherical model
in the hypothesis that all the LMC dark halo consists of compact
lenses. The ellipsoidal model leads to similar results (\cite{mancini}).
A striking feature of the map is the strong near-far
asymmetry, which is not expected, on the contrary, for a self-lensing
population of events.

\section{LMC self-lensing event rate}

We evaluated the microlensing rate in the self-lensing
configuration, i.e. lenses and sources both in the disk and/or in
the bar of LMC. We have taken into account the transverse motion of
the Sun and of the source stars.
We assumed that, to an observer comoving with the LMC center, the
velocity distributions of the source stars and lenses have a
Maxwellian profile, with spherical symmetry.

In the picture of van der Marel et al., within a distance of
about 3 kpc from the center of LMC, the velocity dispersion
(evaluated for carbon stars) along the line of sight can be
considered  constant, $\sigma_{\mathrm{los}}=20.2\pm 0.5$ km/s.
Summing over all fields we find that the expected total number of 
self-lensing events is 
$\sim 1.2$. Clearly, taking also into account the
uncertainties in the parameter used following the van der Marel
model for the LMC the actual number could also be somewhat higher
but hardly more than our upper limit 
estimate of about 3-4 events given in \cite{jetzer02}.

As a further argument, 
assuming  all the 14 events as self-lensing, we
study the  scatter plots  correlating the self-lensing expected
values of some meaningful microlensing variables with the measured
Einstein time or with the self-lensing optical depth. In this way
we can show that a large subset of events is incompatible with
the self-lensing hypothesis.

\begin{figure}
\resizebox{\hsize}{!} {\includegraphics{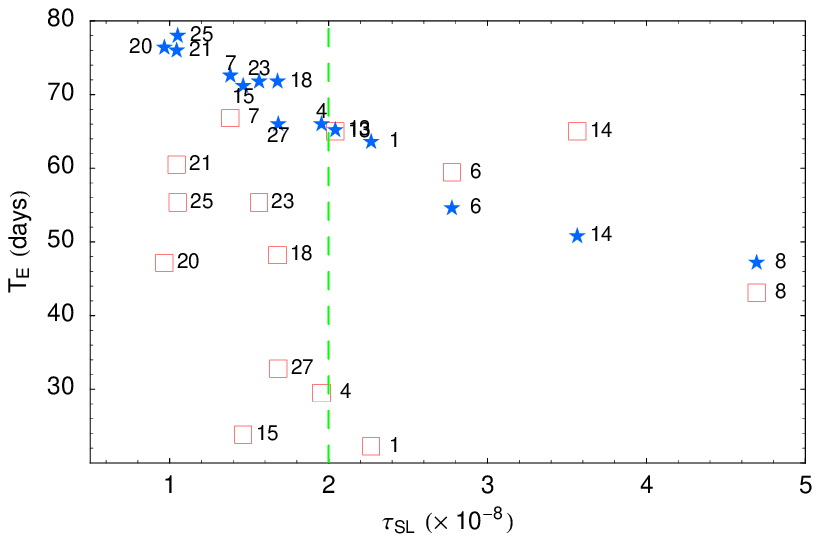}}
{\caption{Scatter plot of the observed (empty boxes) values of the
Einstein time and of the expected values of the median
$T_{\mathrm{E},50\, \%}$ (filled stars), with respect to the
self-lensing optical depth evaluated along the directions of the
events.} \label{tevstau}}
\end{figure}

In Fig. \ref{tevstau} we report on the $y$--axis the observed
values of the Einstein 
time $T_\mathrm{E}$ (empty boxes) as well as the expected
values for self-lensing of the \emph{median}
$T_{\mathrm{E}\,,50\,\%}$ (filled stars) evaluated
\emph{along the directions of the events}. On the $x$--axis we
report the value of the self-lensing optical depth calculated
towards the event position; the optical depth is  growing going
from the outer regions towards the center of LMC according to the
contour lines shown in Fig. \ref{SL}. An interesting feature
emerging clearly is the \emph{decreasing} trend of the  expected
values of the median $T_{\mathrm{E}\,,50\,\%}$, going from the
outside fields with low values of $\tau_{\mathrm{SL}}$ towards the
central fields with higher values of $\tau_{\mathrm{SL}}$. The
variation of the stellar number density and the flaring of the LMC
disk certainly contributes to explain this behaviour.

We now tentatively identify two subsets of events: the nine
falling outside the contour line $\tau_{\mathrm{SL}} = 2 \times
10^{-8}$ of Fig. \ref{SL} and the five falling inside. In the
framework of van der Marel et al. LMC geometry, this  contour line
includes almost fully the LMC bar and two ear shaped inner regions
of the disk, where we expect self-lensing events to be located with
higher probability.

This plot suggests that the self-lensing events have
to be searched among the cluster of events with
$\tau_{\mathrm{SL}}\,>\, 2\times 10^{-8}$, and at the same time
that the cluster of the $9$ events including LMC--1  belongs, very
probably, to a different population.

Moreover, when looking at the spatial distribution of the events
one sees a clear near-far asymmetry in the van der Marel et al.
geometry; they are concentrated along the extension of the bar and
in the south-west side of LMC. Indeed, we have performed a statistical 
analysis of the spatial distribution of the events, which clearly 
shows that the observed asymmetry is greater than the one expected
on the basis of the observational strategy (\cite{mancini}).

Note that, since this study was completed, event MACHO-LMC-23 displayed
a new variation in the EROS2 data (see section 5.4 and Fig.~4).

\section{Pixel lensing towards M31}

\begin{figure}[ht]
\resizebox{\hsize}{!}{\includegraphics{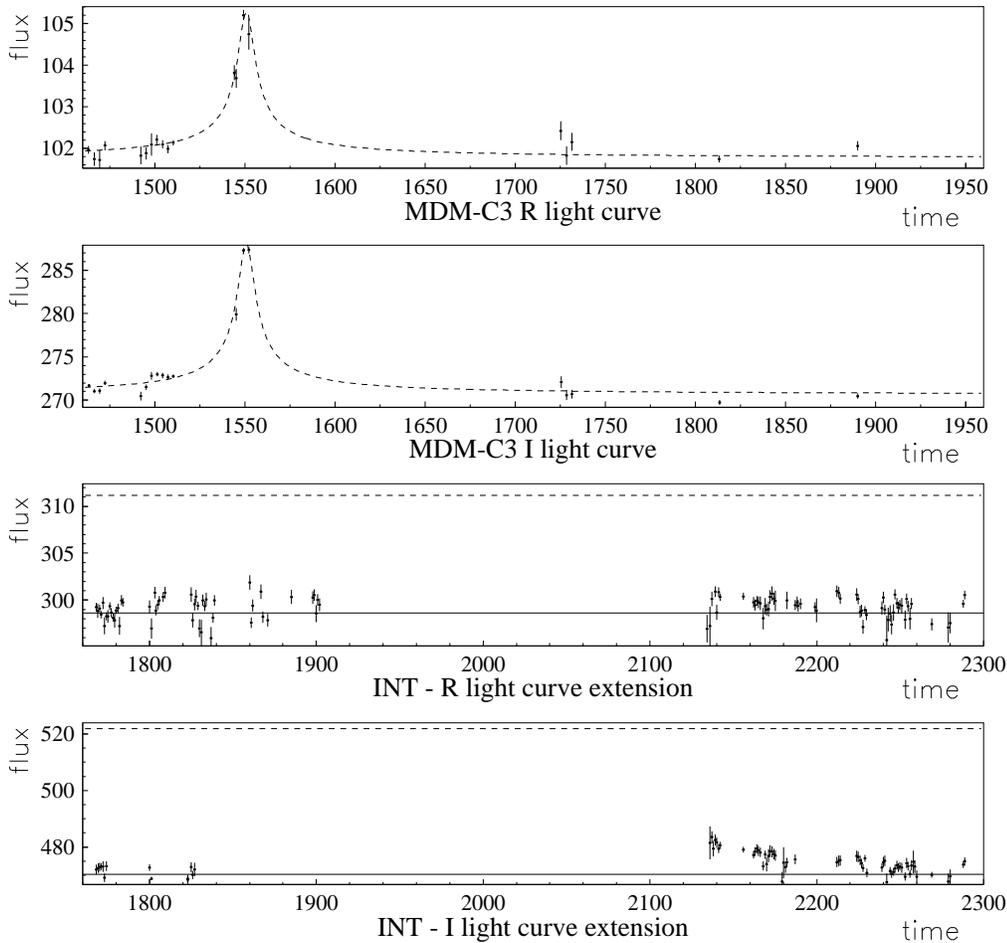}}
{\caption{MDM C3 light curves together with their 
extension into the INT data. 
On the $y$ axis, flux is in ADU/s; on the $x$ axis, 
time is in days, with the origin in J-2449624.5 (both data sets). 
For the MDM light curves the dashed line represent 
the result of the Paczy\'nski fit.
For the INT light curves, shown together with the solid line representing
the baseline is a dashed line representing 
the level of the maximum deviation of flux 
found on the corresponding MDM light curve.}
\label{mdmcl-2}}
\end{figure}

The SLOTT-AGAPE collaboration has been using data collected on the 1.3m 
McGraw-Hill Telescope at the MDM observatory, Kitt Peak (USA).
Observations have been carried out in a two years
campaign, from October 1998 to the end of December 1999.
As a final result of the analysis, taking into account also an extension
of the light curve using INT data, gives three possible candidate events
one of which is reported in Fig. \ref{mdmcl-2}.

The POINT-AGAPE collaboration carried out a survey of M31
by using the Wide Field Camera (WFC) on the 2.5 m INT telescope. 
A first analysis (\cite{point03}) was made with the aim
to detect short ($t_{1/2} < 25$ days) and bright variations
($\Delta R < 21$ at maximum amplification),
compatible with a Paczy\'nski signal.
The first requirement is suggested
by the results on the predicted characteristics
of microlensing events of a Monte Carlo simulation
of the experiment. As an outcome,
seven light curves are detected (\cite{point04}) which can be considered
viable microlensing events.

Once a microlensing event is detected it is important,
given the aim to probe the halo content in form
of macho, to find out its origin, namely, whether it is due to
self-lensing within M31 or to a macho. This is not
straightforward. The spatial distribution of the events
is an important tool, but still unusable given the small statistics.

The light curve of the event 
PA-99-N2 is particularly
interesting because it shows clear deviations
from a Paczy\'nski shape, while remaining
achromatic (and unique) as expected for a microlensing event.
After exploring (\cite{point03a})
different explanations, it is found that the observations
are consistent with an unresolved RGB or AGB star in M31
being microlensed by a binary lens, with
a mass ratio of $\sim 1.2\times 10^{-2}$. 

\section{Results from the final EROS-2 data set}

In a first phase (1990-95), the EROS group (Exp\'erience de Recherche 
d'Objets Sombres) conducted a search 
for low mass machos (where no candidate were found in the mass range 
$10^{-7}$ to $10^{-3}$ solar mass), and a medium sensitivity search 
to higher mass objects ($10^{-4}$ to 1 solar mass). The latter was 
designed to find stellar mass machos, should they be the sole
constituent of the galactic halo. Only two candidates were found,
where seven were expected.
Results from this first phase can be found in \cite{ren97} and 
references therein. 

In order to better study this possible signal, a second survey of the 
Magellanic Clouds was started (EROS-2), with the aim of improving the
sensitivity by at least an order of magnitude.  It was conducted 
between July 1996 and February 2003, using a 1~m dedicated telescope,
the {\sc Marly}, and two wide field 32 million pixel CCD mosaics 
covering 1 deg$^2$.
Details of the setup and operations can be found in \cite{tiss04}
and references therein. We present here the first, preliminary account
of final results using the full Magellanic Clouds data set of EROS-2. 
Interim reports on part of the data 
were presented in \cite{lass00} and \cite{afon03}.

The total number of stars followed was 55 million (including 6 in the 
SMC), compared to 4 million LMC stars in the EROS-1 phase.  The monitored
solid angle is 88 (10)~deg$^2$ in the LMC (SMC).  (For comparison,
the MACHO collaboration monitored 11 million stars over 15~deg$^2$,
mostly in the central region of the LMC (\cite{alcock00a}). We decided 
to restrict our analysis to the 33 million brightest stars in our
sample (29 in the LMC, 4 in the SMC).  In order to analyse a homogeneous
sample, all Magellanic Clouds images were photometered anew, using better
and slightly deeper template images than in our previous reports. 
The typical sampling of the light curves
is about twice per week in each of two passbands. 

We have conducted two independent analyses, aimed at different macho
timescales. When the Einstein timescale is shorter than about two years,
the microlensing light curves display a visible baseline flux. The first
analysis relies on identifying this baseline, and departures from it. 
For longer timescales, because of the six-year duration of the EROS-2
programme, the baseline is not readily seen (or even not at all). 
The second analysis thus relies on light curve shape criteria.

\subsection{Shorter timescale events and their backgrounds}

In the first analysis, we require a single excursion from the baseline, 
that is regular and seen in both passbands; at this stage,
it should loosely agree with the expected microlensing
light curve shape. The selected sample contains mostly variable stars,
but also unphysical variations that originate in PSF variations 
corresponding to optical maintenance of our setup.  We characterize and 
eliminate this spurious signal. The variable stars come mainly in three
categories : ``blue bumpers'', novae and supernovae.  Blue bumpers
are upper main sequence stars that display limited variations (less
than 70\%), that are chromatic (larger in the red passband).
This chromaticity cannot be explained by blending in a microlensing
phenomenon.  Specific cuts deal with this background. Novae and supernovae
have asymmetric light curves, with a faster rise time. Magellanic Clouds 
novae are easily identified and rejected~: 
they are very bright objects with very asymmetric light curves.

After rejection of blue bumpers and novae, supernovae in galaxies 
behind the LMC and SMC represent the dominant population in our sample 
of microlensing candidates at the end of the analysis.
We find 28 such objects, a rate comparable to that found by 
MACHO, see \cite{alcock00a}.  Most of these occur in 
outer regions of the LMC.
Supernovae are
rejected by performing a fit of a function that includes an asymmetry
parameter $S$, and reduces to the usual symmetric microlensing light
curve when $S = 0$, see \cite{tiss04}.
The function is obtained by replacing in the usual microlensing
formula the Einstein timescale $t_E$ with 
$ \, t_E \, [ 1 \, + \, S \arctan ((t-t_0)/t_E) ] $. 
We reject all candidates that have a large asymmetry parameter,
$|S| > 0.3 $.  
In addition, the neighbourhood
of all microlensing candidates is inspected visually on the template 
images. For asymmetric light curves, in more than half of the cases, a galaxy 
or even a cluster of galaxies is visible near the candidate (less than 
30 arcsec). The probability of a chance alignment is very low.  This
reinforces their interpretation as distant supernovae; from 
their flux, we estimate their distance between 500 and 1000~Mpc.

\subsection{SMC results}

The analysis of the SMC data has selected two events. One has been known
already since 1997, see  \cite{alc97b,npd98}. 
The other one is situated in between
the main sequence and the red giant clump, in a thinly populated domain of
the colour-magnitude diagram. The publically available MACHO group data 
on this object display further variation, allowing to reject it.  (Note that
this star lies in the same domain as candidate EROS1-LMC-1, which has
been shown to be a Be variable star, see the discussion below). 
Three long duration candidates had been
detected in the analysis by \cite{afon03} of 5 years of EROS-2 SMC data; 
they did not agree well with a microlensing shape and were thus 
considered dubious. They now
display further variability in the additional 2 years available here;
they have not been selected by our criteria. 

The single EROS-2 
microlensing candidate towards the SMC has a long timescale, 
125~days. If due to a halo microlens, its light curve should display 
a ``microlensing parallax'', i.e. a deformation due to the Earth's
motion around the Sun.  Such a deformation was detected neither in the 
EROS-2 data, nor in the MACHO data. From this, it can be 
concluded that, at 97\% C.L., the lensing object lies in 
the SMC. Thus, the optical depth corresponding to this event
($7 \, 10^{-8}$) is probably best explained by the so-called SMC 
``self-lensing''.

\subsection{LMC results}

The present analysis of the LMC data has selected four new candidates.
(The status of the former EROS LMC microlensing candidates is discussed
in the next section.) One of the candidates, EROS2-LMC-8, 
is a high signal-to-noise microlensing light curve. 
It shows a chromatic variation of a factor
12 (25) in the red (visible) band. Its position in the colour-magnitude
diagram is abnormal, almost
1~mag redder than comparable objects in the main sequence. 
When a possible blending is taken into account
however, the microlensing fit becomes excellent, with agreement
between the timescales in the two bandpasses.  The baseline flux 
corresponding to the magnified object is within the dimmer part of the 
observed LMC main sequence (22.5); this is compatible with the source
being an LMC star. In contrast, the unmagnified flux is now even redder,
and cannot correspond to an LMC star.

We have investigated whether the unmagnified flux could correspond to
the lens being a dim M dwarf in the galactic disk, much in the same way 
as for event MACHO-LMC-5, see \cite{alc01nat}. 
We find a good solution for a 13.5~mag M dwarf (absolute mag)
at a distance of 300~pc. 
A simplified Monte-Carlo simulation shows that the expected
median distance of a disk lens is close to 700~pc, in reasonable
agreement. The observed timescale then corresponds to a transverse 
velocity of 50~km/s, again a normal value.  The median expected 
amplification for such a lens-source configuration is found to be
25, and the observed value is 40. 
Our interpretation of this event could be 
strengthened by better angular resolution images of the source, 
similarly to \cite{alc01nat}. We expect, at the time
of this symposium, that the lens-source separation is 0.15~arcsec.

The three other candidates, EROS2-LMC-9 to 11, have lower signal-to-noise.
The source stars have a baseline flux at about 21st mag and the timescales
range from 35 to 55~days. The magnifications are close to a factor three.
Candidate 10 is in an abnormal region of the colour-magnitude diagram, 0.5
mag redder than comparable objects in the main sequence. It shows the most
chromatic variation of these three candidates, and its rise time (20~d) 
is shorter than its decrease time. For these reasons, we suspect it
may be another distant supernova that our cuts were not
strict enough to reject, possibly the dimmest SN in our analysis. 
Investigation of this candidate is going on.

The analysis described here basically looks for regular variations, such
as those of simple (point source, point lens) microlensing. It is also
able to detect microlensing phenomena with minor deviations from this
simple case, such as ``parallax'' deformations or source size effects.
It may not however be sensitive to caustic-crossing double lens events,
such as MACHO-SMC-98-2, see e.g. \cite{afo00}. For that reason, 
we have performed
another analysis that looks for stars with very fast variations, of at 
least 0.5~mag per day, a typical value in caustic crossing events.
This analysis selected only novae in the Magellanic Clouds. 
No caustic crossing candidate was found.

\subsection{Follow-up of the published microlensing candidates}

Former LMC microlensing candidates have already been retracted after 
having been observed for a longer period of time.  They had displayed a
second variation, that was very unlikely to be due to microlensing of the
second, widely separated component of a double source star; 
the most plausible explanation seemed 
intrinsic stellar variability rather than microlensing.

These retracted candidates are:
EROS1-LMC-2, presented in \cite{aub93} and which displayed
a new variation 8 years later, see \cite{lass00}; 
MACHO-LMC-2
and 3, presented in \cite{alc95} and retracted in \cite{alc97a}; and
candidate EROS2-LMC-4, presented in \cite{lass00} and retracted 
in \cite{milsztajn}.  The fraction of retracted candidates is small,
but non negligible, and no one knows how it will evolve if all
candidates are followed for a much longer period of time. 
We have used our new data set to perform a follow-up of all published
microlensing candidates towards the Magellanic Clouds, both by EROS
and by MACHO. The final EROS2 data provides an additional 
6.6, 3.7 and 5.1 years time base compared to the latest publications by
EROS1, \cite{aub93}, EROS2, \cite{milsztajn} and MACHO, \cite{alcock00a}, 
to check for the stability of the baseline.

There were 5 surviving microlensing candidates from EROS, one from
EROS1 (number 1) and four from EROS2 (numbered 3, 5, 6 and 7). 
Candidate 1
displayed a new variation in 1998, 6.3 years after the first one, 
of similar amplitude (a factor two) and timescale (28~d). 
This second variation is well fit by a microlensing light curve.
Because they are 
separated in time by more than 80 Einstein timescales,  the 
probability that these two bumps correspond to the microlensing of 
a double source star is lower than half a percent. 
This candidate is thus rejected. Moreover, it was
already known to be a Be star (\cite{beau95})
and was thus suspected of being variable.
Candidate 3 also displayed new, irregular variations between 
1999 and 2002, and was thus rejected.

The light curves of candidates 5 to 7 have been improved, due to the 
better template images in the present analysis. This made 
apparent an asymmetry in rise and fall times, that allowed
to identify them as supernovae. Note that there were no cut targeted
at rejecting distant supernovae in previous EROS analyses, and that
our candidate 5 is identical to MACHO candidate 26, which was 
rejected by \cite{alcock00a} for the very same reason.

The conclusion is that {\it none} of the former EROS microlensing 
candidates are still considered valid.

We have also attempted to follow all 13 MACHO candidates selected by
their analysis A (stricter cuts). We were able to identify unambiguously
nine of them. Our data provide 5.1 (resp. 3.1) additional years compared 
to the published (resp. Web available) light curves. One star, 
MACHO-LMC-23, displays a new variation in Dec~2001, very similar 
to that observed by the MACHO group in Feb~1995 (variation of 1 mag and
timescale of 40~days). The corresponding light curve is shown in 
Fig.~\ref{macho23}.  Note that this variation is compatible with 
being achromatic, and is also well fit by a microlensing light curve. 

\begin{figure}[ht]
\centering
{\caption{The light curves of candidate MACHO-LMC-23, as seen by EROS2 
between 1996 and 2003 (top : red passband; bottom : visible). 
The flux is in arbitrary units; 
the time is in days, with the origin at JD~=~2450000. 
The bump at day 2250 is a new variation, 6.8 years after that  
originally seen by the MACHO group, before the startup of EROS2.
}
\label{macho23}}
\end{figure}

\subsection{Limits on the macho content of the halo}

This search for microlensing phenomena with timescales shorter than
2 years in the complete EROS2 Magellanic Clouds data set has selected
four LMC candidates and one SMC candidate. The two high 
signal-to-noise events are most likely microlensing
due to an SMC lens (the SMC candidate) 
and a galactic disk lens (event 8 towards the LMC).
The remaining 3 events, if interpreted as machos, correspond to an 
optical depth of $1.5 \, 10^{-8}$, i.e. 3\% of the standard spherical 
halo. Taking a more conservative approach, one can choose to present
a 95\% C.L. upper limit on the macho content of the halo, based on
these three candidates.  A {\it preliminary} version of this limit 
is shown in Fig.~\ref{limit04}. 
It uses {\it only} the EROS2 LMC data set; combination with 
both the EROS1 and the EROS2 SMC results will be done later.
The limit is better than 12\% of the halo for macho masses 
between 0.0002 and 1 solar mass.

This limit is compared to the previous EROS2 LMC result of \cite{lass00},
to show the increase in sensitivity since 2000.  It is also compared
to the signal presented by \cite{alcock00a}. Our result is compatible
with the lowest part of the domain allowed by the MACHO group analysis. 
It must be recalled however that the fields monitored by MACHO
and EROS2 are not identical. A better comparison is in progress; 
in particular, we are trying to evaluate whether the EROS2 result is 
compatible with the central value of \cite{alcock00a}, i.e.
a 20\% macho component of the halo.  
 
\begin{figure}[ht]
{\caption{The {\it preliminary}
95\% C.L. upper limit on the macho content of the halo 
obtained from the final EROS2 LMC data only  
(red curve between 0.0001 and 100 
solar masses). The lower cross is the value of the optical depth 
towards the LMC corresponding to the three candidates. The dashed 
line that goes through this cross shows what the limit would be  
in the absence of microlensing candidates.  This limit is compared 
to that presented in \cite{lass00} using the same method, but all 
EROS data sets available at the time (green curve between $10^{-7}$ 
and 10 solar mass). The smaller orange-shaded contour  between 0.1 and 1 
solar mass is the signal presented by \cite{alcock00a}, 
with the lighter cross indicating their preferred solution. 
}
\label{limit04}}
\end{figure}

Finally, we remark that the present value of the LMC optical depth from 
EROS2 seems to be compatible with ``self-lensing'' within the LMC.

\subsection{Longer timescale events}

We have also searched for events with timescales longer than two years
in the EROS2 data, using only light curve shape criteria.
This analysis is still in progress. At present, 10 candidates have
been selected. All but one are located in the central fields of 
the LMC, that contain 30\% of the stars monitored by EROS2.  
This in itself shows that they cannot all be microlensing phenomena
caused by halo objects, as such events should follow the LMC stellar
distribution.  No candidate was selected towards the SMC.

Nine of these 10 events are found in the MACHO publically available
light curve database.  The combination of the MACHO and EROS2 data
indicate that these nine events cannot be caused by microlensing.
When this analysis is completed, the microlensing sensitivity to
high mass objects will reach the vicinity of 1000~solar mass, thus 
bridging the gap with the recent results on higher mass machos
from \cite{yoo04},
obtained from a study of loosely bound double stars of the galactic 
halo.

\section{Conclusions} 

We have presented the results of the computation of the
optical depth for microlensing 
towards the LMC using a recent picture 
of the LMC disk. An interesting feature which emerges 
is a near--far asymmetry of the optical depth for 
lenses located in the LMC halo, which is not the case
for self-lensing.
Furthermore, we showed 
that the timescale distribution of 
the events and their spatial variation across the
LMC disk offers possibilities of identifying the dominant lens
population. 
Through this analysis we have been able
to identify a large subset of events that
can not be accounted for by self-lensing.

As a general outcome of presently available
pixel lensing results, we can clearly infer that
the detection of microlensing events towards M31 is now established.
The open issue to be still explored is the 
study of the M31 halo fraction in form of machos.

Finally, the first account of the analysis of the full EROS2 data set
towards the Magellanic Clouds was presented. This data set has presently
the largest sensitivity to halo machos. The small number of microlensing
candidates is lower than expected from the results of the MACHO group.
From this, one derives strict limits on the abundance of machos. 
A few microlensing candidates have been shown to vary again, 
many years later; they should be removed from the candidates census.

------------------------------------------------------------------

\end{document}